\documentclass[pra,a4paper,aps,onecolumn,showpacs,superscriptaddress,groupedaddress]{revtex4}
\usepackage[dvips]{graphicx}
\usepackage{ae}
\usepackage[T1]{fontenc}
\usepackage[ansinew]{inputenc}
\usepackage{amsmath}
\usepackage{amssymb}
\usepackage{graphicx}
\usepackage{color}
\usepackage[colorlinks]{hyperref}
\usepackage{lscape}
\begin{document}
\title{Monogamy of Measurement-Induced NonLocality}
\author{Ajoy Sen}
\email{ajoy.sn@gmail.com}
\affiliation{Department of Applied Mathematics, University of Calcutta, 92, A.P.C. Road, Kolkata-700009, India.}
\author{Debasis Sarkar}
\email{dsappmath@caluniv.ac.in}
\affiliation{Department of Applied Mathematics, University of Calcutta, 92, A.P.C. Road, Kolkata-700009, India.}
\author{Amit Bhar}
\email{bhar.amit@yahoo.com}
\affiliation{Department of Mathematics, Jogesh Chandra Chaudhuri College, 30, Prince Anwar Shah Road, Kolkata-700033, India.}

\begin{abstract}
Measurement-Induced NonLocality was introduced by Luo and Fu (Phys. Rev. Lett. \textbf{106}, 120401,(2011)) as a measure of nonlocality in a bipartite state. In this paper we will discuss monogamy property of measurement-induced nonlocality for some three- and four-qubit classes of states. Unlike discord, we find quite surprising results in this situation. Both the GHZ and W states satisfy monogamy relations in the three-qubit case, however, in general there are violations of monogamy relations in both the GHZ-class and W-class states. In case of four-qubit system, monogamy holds for most of the states in the generic class. Four qubit GHZ does not satisfy monogamy relation, but W-state does. We provide several numerical results including counterexamples regarding monogamy nature of measurement induced nonlocality. We will also extend our results of generalized W-class to n-qubit.
\end{abstract}
\date{\today}
\maketitle

\section{INTRODUCTION}

Nonlocality is in the heart of quantum world. From Bell's theorem\cite{bell}, it is understood that no local hidden variable theory could replace quantum theory as a theory of physical world. The existence of entangled states in composite quantum systems assures the nonlocal behavior of quantum theory. Generally, violation of Bell's inequality is taken as the signature of nonlocality. Entangled states play an important role to show the violation of Bell's inequality. However nonlocality can be viewed from other perspectives. For example, there exist sets of locally indistinguishable orthogonal pure product states\cite{bennett}, used to show nonlocality without entanglement. Recently introduced measurement-induced nonlocality\cite{luo} (in short, MIN), is one of the ways to detect nonlocality in quantum states by some locally invariant measurements. Locally invariant measurements can not affect global states in classical theory but this is possible in quantum theory. So MIN is a type of nonlocal correlation which can only exists in quantum domain. MIN is defined in such a way that it is non-negative for all states, invariant under local unitary and vanishes on product state. In this sense, MIN could be observed as a type of nonlocal correlation which is induced by certain measurement. Although it is induced by some measurement, it is not a measure of quantum correlation in true sense. But it can be treated as a measure of nonlocality, induced by some kind of measurement.

Now a natural question arises about the shareability of quantum correlations in multipartite states. It may be monogamous or may be polygamous. It is known that entanglement is monogamous\cite{wootters}. In this work we will investigate the monogamous behavior of MIN. We will check the monogamy property for some classes of states in both three and four-qubit system. Unlike discord\cite{giorgi,prabhu,bruss}, here we will show that both the three-qubit generalized GHZ and W states satisfy monogamy relations, however there is violations of monogamy relation if we consider the generic whole class of pure three-qubit states. In the case of pure four-qubit system we consider the most important generic class of states. It contains usual GHZ, maximally entangled states in the sense of Gour et.al.\cite{gour}. Most of the states satisfy monogamy relations. There are two important subclasses of the generic class, say, $\mathcal{M}$ and $\tau_{min}$. In one subclass monogamy holds but in another it does not. In particular, GHZ state violates monogamy relation and W state satisfies it. Therefore, monogamous relations of MIN are quite different from that of some important measures of correlation\cite{dakic,zurek} and it acts as distinguishing feature of some class of states. Our paper is organized as follows - in section II we will review some of the basic properties of MIN. In section III we will explain and discuss some four-qubit classes of states which we will require in our work. Section IV is devoted to the notion of monogamy for MIN. Section V contains results on pure three qubit systems and section VI contains results on four-qubit system. Several counterexamples and numerical figures are discussed in both the above sections.

\section{OVERVIEW ON MIN}
Let $\rho$ be any bipartite state shared between two parties A and B. Then MIN (denoted by $N(\rho)$) is defined as\cite{luo},
\begin{equation}
N(\rho):=\max_{\Pi^{A}}\parallel \rho-\Pi^{A}(\rho)\parallel^2
\end{equation}
where maximum is over all von Neumann measurements $\Pi^{A}$ which do not disturb $\rho_{A},$ the local density matrix  of A, i.e., $\Sigma_{k}\Pi_{k}^{A}\rho_{A}\Pi_{k}^{A}=\rho_{A}$ and $\|.\|$ is taken as the Hilbert Schmidt norm (i.e. $\parallel X \parallel=[Tr(X^{\dag}X)]^{\frac{1}{2}}$). It is in some sense, dual to that of geometric measure of discord. Physically, MIN quantifies the global effect caused by locally invariant measurements. MIN has applications in general dense coding, quantum state steering etc. MIN vanishes for product state and remains positive for entangled states. For pure states MIN reduces to linear entropy like geometric discord\cite{dakic}. Explicit formula of MIN for  $ 2 \otimes{n} $ system, $ m \otimes{n} $ (if $\rho_{A}$ is non-degenerate) system and an explicit upper bound for $ m \otimes{n} $ system were obtained by Luo and Fu in \cite{luo}. Later Mirafzali \textit{et.al.} \cite{mirafzali} formulate a way to reduce the problem of degeneracy in $ m \otimes{n} $ system and evaluate it for $ 3 \otimes{n} $ dimensional systems. MIN is invariant under local unitary, i.e., in true sense, it is a nonlocal correlation measure. The set of all zero MIN states is non-convex. Guo and Hou \cite{guo} derived the conditions for the nullity of MIN. They have found the set of states with zero MIN is a proper subset of the set of all classical-quantum states, i.e., zero discord states. MIN for classical-quantum states vanishes if each eigen-subspace of $\rho_{A}$ is one dimensional. It therefore reveals that non-commutativity is the cause of this kind of nonlocality in quantum states. Recently, in \cite{xi}, MIN has been quantified in terms of relative entropy to give it another physical interpretation. However, in our work we have used the original definition of MIN.

Suppose $H^{A}$, $H^{B}$ are the Hilbert spaces associated with parties A and B respectively and $L(H^{A})$, $L(H^{B})$ denote the Hilbert space of linear operators acting on $H^{A}$, $H^{B}$ with the inner product defined by $\langle X|Y\rangle:=trX^{\dag}Y $. We state two important results which we will use in our work. \\

{\textbf{Theorem 1}:(Luo and Fu \cite{luo}) \textit{Let $|\psi\rangle_{AB}$ be any bipartite pure state with Schmidt decomposition  $|\psi\rangle_{AB}=\sum_{i}\sqrt{s_{i}}|\alpha_{i}\rangle_{A}|\beta_{i}\rangle_{B},$ then $N(|\psi\rangle_{AB})=1-\sum_{i}s_{i}^{2}$}.\\\\

\textbf{Theorem 2}:(Luo and Fu \cite{luo}) \textit{Let $\rho_{AB}$ be any state of $2\otimes{n}$ dimensional system written in the form
\begin{equation}
\begin{split}
\rho_{AB}=\frac{1}{\sqrt{2n}}\frac{I^{A}}{\sqrt{2}}\otimes{\frac{I^{B}}{\sqrt{n}}}+\sum_{i=1}^{3}x_{i}X_{i}^{A}\otimes{\frac{I^{B}}{\sqrt{n}}}
+\frac{I^{A}}{\sqrt{2}}\otimes{\sum_{i=1}^{n^{2}-1}y_{i}Y_{i}^{B}}+\sum_{i=1}^{3}\sum_{j=1}^{n^{2}-1}t_{ij}X_{i}^{A}Y_{j}^{B}
\end{split}
\end{equation}
where $\{X_{i}^{A}:\,$ i=0,1,2,3\} and $\{Y_{j}^{B}:\,{ j=0,1,2,...,n^{2}-1}\}$ are the orthonormal Hermitian operator bases for $L(H^{A})$ and $L(H^{B})$ respectively with $X_{0}^{A}=I^{A}/\sqrt{2},Y_{0}^{B}=I^{B}/\sqrt{n} $. Then
\begin{equation}
\begin{split}
N(\rho_{AB})&=tr TT^{t}-\frac{1}{\|\textbf{x}\|^{2}}\textbf{x}^{t}TT^{t}\textbf{x}\qquad{if\textbf{\,x}\neq{\textbf{0}}}\\
&=trTT^{t}-\lambda_{3}\,\,\,\quad\quad\qquad\qquad{if\textbf{\,x}=\textbf{0}}
\end{split}
\end{equation}
where the matrix $T=(t_{ij})$ with $\lambda_{3}$ being minimum eigenvalue of $TT^{t}$ and $\|\textbf{x}\|^{2}:=\sum_{i}x_{i}^{2}$ with $\textbf{x}=(x_{1},x_{2},x_{3})^{t}$
}

Before going to discuss the monogamy properties of measurement-induced nonlocality, specifically for three- and four-qubit systems, we first mention some important classes of states in four qubit systems with some discussions on their entanglement behavior.

\section{SOME SPECIAL FOUR-QUBITS CLASSES}
Four qubit pure states can be classified into nine groups\cite{moor}. Among them the generic class is given by
\begin{equation}
\mathcal{A}\equiv{\{\sum_{j=0}^{3}z_{j}u_{j}:\sum_{j=0}^{3}|z_{j}|^{2}=1, z_{i}\in{\mathbb{C}},i=0,1,2,3\}}
\end{equation}
where $u_{0}\equiv{|\phi^{+}\rangle|\phi^{+}\rangle}$, $u_{1}\equiv{|\phi^{-}\rangle|\phi^{-}\rangle}$, $u_{2}\equiv{|\psi^{+}\rangle|\psi^{+}\rangle}$ and $|\phi^{\pm}\rangle=(|00\rangle\pm|11\rangle)/\surd{2}$, $|\psi^{\pm}\rangle=(|01\rangle\pm|10\rangle)/\surd{2}$. Consider two important subclasses $\mathcal{M}$ and $\tau_{min}$ of the generic class $\mathcal{A}$ which are defined as,
\begin{equation}
\mathcal{M}=\{\sum_{j=0}^{3}z_{j}u_{j}: \sum_{j=0}^{3}|z_{j}|^{2}=1, \sum_{j=0}^{3}z_{j}^{2}=0\}
\end{equation}
and
\begin{equation}
\tau_{min}=\{\sum_{j=0}^{3}x_{j}u_{j}: \sum_{j=0}^{3}x_{j}^{2}=1, x_{j}\in{\mathbb{R},j=0,1,2,3}\}
\end{equation}
These two subclasses are important in the sense that $\mathcal{M}$ is the maximally entangled class and $\tau_{min}$ has least amount of bipartite entanglement according to the definition of maximally entangled states given by Gour \textit{et. al.} \cite{gour}. Consider a pure bipartite state $|\psi_{AB}\rangle\in\mathbb{C}^{m}\otimes{\mathbb{C}^{n}}$. Then the tangle is defined as
\begin{equation}
\tau_{AB}=2(1-tr\rho_{A}^{2})
\end{equation}
where $\rho_{A}=Tr_{B}|\psi_{AB}\rangle\langle\psi_{AB}|$. Now for a pure state $|\psi_{ABCD}\rangle$, shared between four parties, three quantities $\tau_{1},\tau_{2},\tau_{ABCD}$ are defined as
\begin{equation}
\tau_{1}\equiv{\frac{1}{4}(\tau_{A|BCD}+\tau_{B|ACD}+\tau_{C|ABD}+\tau_{D|ABC})}
\end{equation}
\begin{equation}
\tau_{2}\equiv{\frac{1}{3}(\tau_{AB|CD}+\tau_{CA|BD}+\tau_{DA|BC})}
\end{equation}
\begin{equation}
\tau_{ABCD}=4\tau_{1}-3\tau_{2}
\end{equation}
For the above two subclasses $\mathcal{M}$ and $\tau_{min}$ we have $\tau_{1}(\mathcal{M})=1$, $\tau_{2}(\mathcal{M})=\frac{4}{3}$, $\tau_{ABCD}(\mathcal{M})=0$ and $\tau_{1}(\tau_{min})=1$, $\tau_{2}(\tau_{min})=1$, $\tau_{ABCD}(\tau_{min})=1$.
Four-qubit GHZ state belongs to the class $\tau_{min}$.

\section{MONOGAMY}
Monogamy is an important aspect in our physical world which restricts the shareability of bipartite correlation. Entanglement is an example of quantum correlation which is monogamous w.r.t. the tangle. Mathematically a correlation measure $Q$ is said to be monogamous iff for any n-party state $\rho_{A_{1}A_{2}...A_{n}}$ the relation
\begin{equation}
\sum_{k=1,k\neq{i}}^{n}Q(\rho_{A_{i}A_{k}})\leq{Q(\rho_{A_{i}|A_{1}A_{2}...A_{i-1}A_{i+1}...A_{n}})}
\end{equation}
holds for all $i=1,2,...,n$.
Now consider an n-party state $\rho_{12...n}$. Let the locally invariant measurement be done on the party 1. Then MIN is defined as $N(\rho_{1|2...n})=\parallel \rho_{12...n}-\Pi_{1}^{*}(\rho_{1|2...n})\parallel^{2}$, where $\Pi_{1}^{*}=\{\pi_{1k}^{*}\}$ is the optimal measurement done by the party 1 which does not change its local density matrix, i.e., $\rho_{1}=\Sigma_{k}\pi_{1k}^{*}\rho_{1}\pi_{1k}^{*}$. On the other hand, since $\rho_{1}=Tr_{2,3,...,n}(\rho_{12...n})=Tr_{j}(Tr\rho_{1j}), j=2,3,...,n$, the optimal measurement also does not change the local density matrices for all two-party reduced states of $\rho_{12...n}$ of the kind, $\rho_{1j}=Tr_{2,3,...,j-1,j+1,...,n}(\rho_{12...n})$. Then $\Sigma_{j}N(\rho_{1j})\geq{\sum_{j}\|\rho_{1j}-\Pi_{1}^{*}(\rho_{1j})\|^{2}}$. So in case of polygamy $N(\rho_{1|2...n})<\sum_{j}\|\rho_{1j}-\Pi_{1}^{*}(\rho_{1j})\|^{2}$ and if $N(\rho_{1|2...n})\geq\sum_{j}\|\rho_{1j}-\Pi_{1}^{*}(\rho_{1j})\|^{2}$ then the state is monogamous w.r.t. MIN.

\section{MONOGAMY IN THREE QUBIT SYSTEM}
Any three qubit pure state $|\psi_{ABC}\rangle$ has a generic form $\lambda_{0}|000\rangle+\lambda_{1}e^{i\theta}|100\rangle+\lambda_{2}|101\rangle+\lambda_{3}|110\rangle+\lambda_{4}|111\rangle$, where $\lambda_{i}\in{\mathbb{R}},\theta\in{[0,\pi]},\sum_{i}\lambda_{i}^{2}=1 $\cite{acin}. This class includes GHZ class. W class can also be availed by  putting $\lambda_{4}=0$. For the general state $|\psi_{ABC}\rangle$, the reduced density matrix $\rho_{AB}=Tr_{C}|\psi_{ABC}\rangle\langle\psi_{ABC}|$ has the form
\[
 \rho_{AB}=
 \begin{bmatrix}
  \lambda_{0}^{2} & 0 & \lambda_{0}\lambda_{1}e^{-i\theta} & \lambda_{0}\lambda_{3}\\
  0 & 0 & 0 & 0 \\
  \lambda_{0}\lambda_{1}e^{i\theta} & 0& \lambda_{1}^{2}+\lambda_{2}^{2} & \lambda_{1}\lambda_{3}e^{i\theta}+\lambda_{2}\lambda_{4} \\
  \lambda_{0}\lambda_{3} & 0 & \lambda_{1}\lambda_{3}e^{-i\theta}+\lambda_{2}\lambda_{4} &  \lambda_{3}^{2}+\lambda_{4}^{2}
 \end{bmatrix}
\]
 where the correlation matrix $ T=(t_{ij})$ is obtained from the relation $t_{ij}=tr (\rho \frac{\sigma_{i}}{\sqrt{2}}\otimes{\frac{\sigma_{j}}{\sqrt{2}}}),i,j=1,2,3; \sigma_{i}$'s being the Pauli matrices:
\[
 T=
 \begin{bmatrix}
  \lambda_{0}\lambda_{3} & 0 & \lambda_{0}\lambda_{1}\cos{\theta} \\
  0 & -\lambda_{0}\lambda_{3} & -\lambda_{0}\lambda_{1}\sin{\theta} \\
  -\lambda_{1}\lambda_{3}\cos{\theta}-\lambda_{2}\lambda_{4} & -\lambda_{1}\lambda_{3}\sin{\theta}& 0.5-\lambda_{1}^{2}-\lambda_{3}^{2} \\
 \end{bmatrix}
\]
Other reduced density matrix $\rho_{AC}$ and its corresponding correlation matrix could be written from the expressions of $\rho_{AB}$ and $T$ by only interchanging $\lambda_{2}$ and $\lambda_{3}$. Coherent vector for both the reduced density matrices is  $\textbf{x}=(\lambda_{0}\lambda_{1}\cos{\theta},-\lambda_{0}\lambda_{1}\sin{\theta},\lambda_{0}^{2}-0.5)^{t}$.\\
Clearly, $\parallel \textbf{x} \parallel=0$ iff $\lambda_{0}^{2}=0.5,\lambda_{1}^{2}=0$.  In case of $\parallel \textbf{x} \parallel=0$ we have,
\begin{eqnarray}
N(\rho_{AB})&=&2a+c-\min\{a,\frac{1}{2}(a+c-\sqrt{(a-c)^{2}+4b^{2}})\}\\
N(\rho_{AC})&=&2g+k-\min\{g,\frac{1}{2}(g+k-\sqrt{(g-k)^{2}+4b^{2}})\}
\end{eqnarray}
\begin{eqnarray}
N(\rho_{A|BC})&=&0.5
\end{eqnarray}
where
\begin{eqnarray}
   a&=&\lambda_{0}^{2}\lambda_{3}^{2}  \\
   b&=&-\lambda_{0}\lambda_{2}\lambda_{3}\lambda_{4} \\
   c&=&\lambda_{2}^{2}\lambda_{4}^{2}+(0.5-\lambda_{3}^{2})^{2} \\
   g&=&\lambda_{0}^{2}\lambda_{2}^{2} \\
   k&=&\lambda_{3}^{2}\lambda_{4}^{2}+(0.5-\lambda_{2}^{2})^{2}
\end{eqnarray}
Now, the minimum value of $N(\rho_{AB})+N(\rho_{AC})$ is 0.5. So in case of $ \parallel \textbf{x} \parallel=0$ monogamy is violated for most of the states. For example, we can consider a state with $\lambda_{2}=\lambda_{3}=\lambda_{4}=\sqrt{\frac{1}{6}}$. In this case $N(\rho_{AB})+N(\rho_{AC})=0.516046>0.5$. Numerical simulation of $10^{6}$ random states (the states are generated by choosing random  $\lambda_{i}$'s $,i=0,1,...,5$ with uniform distribution) shows that around $0.02\% $ GHZ class states and around $ 20\% $ W class states satisfy equality of the monogamy relation.\\
When $\parallel \textbf{x} \parallel\neq 0$ we have
\begin{eqnarray}
N(\rho_{A|BC})&=2\lambda_{0}^{2}(\lambda_{2}^{2}+\lambda_{3}^{2}+\lambda_{4}^{2})
\end{eqnarray}
\begin{eqnarray}
N(\rho_{AB})&=a+b+c-\frac{1}{\|\textbf{x}\|^{2}}\textbf{x}^{t}TT^{t}\textbf{x}\\
N(\rho_{AC})&=g+f+k-\frac{1}{\|\textbf{x}\|^{2}}\textbf{x}^{t}TT^{t}\textbf{x}
\end{eqnarray}
where
\begin{eqnarray}
a&=&\lambda_{0}^{2}\lambda_{3}^{2}+\lambda_{0}^{2}\lambda_{1}^{2}\cos^{2}\theta \\
b&=&\lambda_{0}^{2}\lambda_{3}^{2}+\lambda_{0}^{2}\lambda_{1}^{2}\sin^{2}\theta \\ c&=&(\lambda_{2}\lambda_{4}+\lambda_{1}\lambda_{3}\cos\theta)^{2}+\lambda_{1}^{2}\lambda_{3}^{2}\sin^{2}\theta+(0.5-\lambda_{1}^{2}-\lambda_{3}^{2})^{2}\\
g&=&\lambda_{0}^{2}\lambda_{2}^{2}+\lambda_{0}^{2}\lambda_{1}^{2}\cos^{2}\theta\\
f&=&\lambda_{0}^{2}\lambda_{2}^{2}+\lambda_{0}^{2}\lambda_{1}^{2}\sin^{2}\theta\\
k&=&(\lambda_{3}\lambda_{4}+\lambda_{1}\lambda_{2}\cos\theta)^{2}+\lambda_{1}^{2}\lambda_{2}^{2}\sin^{2}\theta+(0.5-\lambda_{1}^{2}-\lambda_{2}^{2})^{2}
\end{eqnarray}
Now the maximum value of $ N(\rho_{AB})+N(\rho_{AC})$ is 0.5. Hence three qubit pure states with $\parallel \textbf{x} \parallel\neq 0$ satisfies the monogamy relation.
Specifically, the three qubit generalized GHZ class of pure states($\alpha|000\rangle +\beta|111\rangle$) is monogamous in the region $\alpha\neq \beta$ and the monogamy relation holds good with equality if $\alpha= \beta(=\frac{1}{\sqrt{2}})$. In the three qubit generalized W-Class states($\alpha|001\rangle +\beta|010\rangle+\gamma|001\rangle$) we have $N(\rho_{AB})+N(\rho_{AC})=N(\rho_{A|BC})=2|\alpha|^{2}(1-|\alpha|^{2})$. That is, the monogamy relation holds with equality.

\section{MONOGAMY IN FOUR QUBIT SYSTEM}

Consider a four-qubit generic pure state $|\psi_{ABCD}\rangle$ shared between four parties $A, B, C,D$ i.e., from the class $\mathcal{A}$ where
\begin{equation}
|\psi_{ABCD}\rangle=\sum_{j=0}^{3}z_{j}u_{j}, \quad{\sum_{j=0}^{3}|z_{j}|^{2}=1}.
\end{equation}
We consider the two-qubit reduced density matrices $\rho_{AB}, \rho_{AC}$ and $\rho_{AD}$ of $\rho=|\psi_{ABCD}\rangle\langle\psi_{ABCD}|$. Each reduced density matrix is of the form

\[
\frac{1}{4}
 \begin{bmatrix}
  \alpha & 0 & 0 & \beta\\
  0 & \gamma & \delta & 0 \\
  0 & \delta & \gamma & 0 \\
  \beta & 0 & 0 & \alpha
 \end{bmatrix}
\]
where $\alpha,\beta,\gamma,\delta$ are some suitable functions of  $z_{0},z_{1},z_{2},z_{3}$ such that $\alpha,\beta,\gamma,\delta\in{\mathbb{R}}$ and $\alpha+\gamma=2$. We define 4 quantities $a=z_{0}+z_{1},b=z_{0}-z_{1},c=z_{2}+z_{3},d=z_{2}-z_{3}$.\\
Then for $\rho_{AB}$:
\begin{eqnarray}
  \alpha&=&2(|z_{0}|^{2}+|z_{1}|^{2}) \\
  \beta&=&2(|z_{0}|^{2}-|z_{1}|^{2}) \\
  \gamma&=&2(|z_{2}|^{2}+|z_{3}|^{2}) \\
  \delta&=&2(|z_{2}|^{2}-|z_{3}|^{2})
\end{eqnarray}
for $\rho_{AC}$:
\begin{eqnarray}
  \alpha&=&(|a|^{2}+|c|^{2}) \\
  \beta&=&2 Re(\overline{a}c) \\
  \gamma&=&(|b|^{2}+|d|^{2}) \\
  \delta&=&2 Re(\overline{b}d)
\end{eqnarray}
and for $\rho_{AD}$:
\begin{eqnarray}
  \alpha&=&(|a|^{2}+|d|^{2}) \\
  \beta&=&2 Re(\overline{a}d) \\
  \gamma&=&(|b|^{2}+|c|^{2}) \\
  \delta&=&2 Re(\overline{b}c)
\end{eqnarray}
The elements of the correlation matrix $T$ can be obtained from $t_{ij}=tr(\rho_{AX} X_{i}\otimes{Y_{j}}), ~X=B,C,D$. Eigenvalues of the matrix $TT^{t}$ are of the form $k(\beta\pm\delta)^{2},k(\alpha-\gamma)^2$ with $k=\frac{1}{16}$. The coherent vector $\textbf{x}=(x_{1},x_{2},x_{3})^{t}$ as obtained from the relation $x_{i}=tr(\rho_{AX} X_{i}\otimes{I}), ~X=B,C,D$ is zero for all the three cases. Hence we have(by Theorem 2),
\begin{eqnarray}
N(\rho_{AX})&=&k[2(\beta^{2}+\delta^{2})+(\alpha-\gamma)^{2}-\lambda_{3}]\quad \text{where}\quad X=B,C,D\\
\lambda_{3}&=&min\{(\beta+\delta)^{2},(\alpha-\gamma)^2,(\beta-\delta)^{2}\}
\end{eqnarray}
On the other hand, we can write the state $\rho_{A|BCD}$ in the form $\frac{1}{\sqrt{2}}(|0\phi_{0}\rangle+|1\phi_{1}\rangle)$ where $|\phi_{0}\rangle, |\phi_{1}\rangle$ are mutually orthonormal. Since this is a pure state we have $N(\rho_{A|BCD})=0.5$ (using Theorem 1). Numerical simulation for $10^{6}$ generic states shows that about 66\% of them satisfies monogamy relation. Hence in general four-qubit generic class is not monogamous w.r.t. MIN. So there exist quantum states whose locally shared nonlocality exceeds the amount of globally shared nonlocality.\\
\begin{figure}[ht]
\centering
\scalebox{0.5}
{\includegraphics{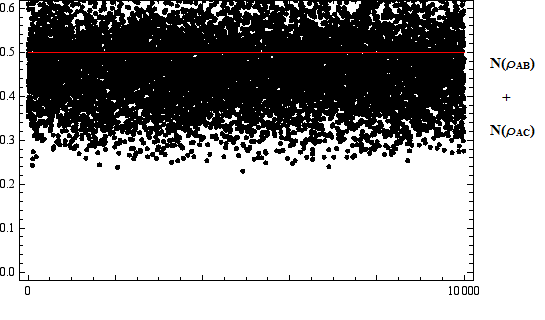}}
\\ \label{fig:1}
\caption{\textit{$10^{5}$ random simulation for the states, generated by choosing random  $\lambda_{i}$'s $,i=0,1,2,3$ with uniform distribution shows about 66\% violation of monogamy for the generic class $\mathcal{A}$. The red line marks the departure from monogamy.}}
\end{figure}\\
However for the sub classes $\mathcal{M}$ and $\tau_{min}$ we can still check whether monogamy relation holds or not, and if not then what is the amount of violation. These will be illustrated in the next few results.\\\\
\textbf{Theorem 3}: \textit{Consider a four-qubit system. Then for any pure state $\rho$, belonging to the generic class $\mathcal{A}$ we have $N(\rho_{XY})\leq\sum_{i=1}^{3}\lambda_{i}$ where $\rho_{XY}$ denotes any bipartite reduced density matrix of $\rho$ and $\lambda_{i}$'s are the eigenvalues of $TT^{t}$, T being the correlation matrix of $\rho_{XY}$}.\\\\
\textit{Proof}: The proof of the theorem is very easy as it directly follows form Theorem 2 keeping in mind that the eigenvalues of $TT^{t}$} are all positive and the Bloch vector $\|\textbf{x}\|=\textbf{0}$ for all bipartite reduced states of any generic state.\\ We will use this theorem in our proof of the the next two results. $\blacksquare$ \\\\
\textbf{Theorem 4}: \textit{Let $\rho_{ABCD}$ be any four-qubit pure state belonging to the class $\mathcal{M}$. Then
$N(\rho_{AB})+N(\rho_{AC})+N(\rho_{AD})\leq \frac{1}{4}$ }\\\\
\textit{Proof}: Let us consider any four qubit pure state $\rho_{ABCD}=|\psi\rangle_{ABCD}\langle\psi|$ belonging to the class $\mathcal{M}$ i.e., $|\psi\rangle_{ABCD}=\sum_{j=0}^{3}z_{j}u_{j}$, with $\sum_{j=0}^{3}|z_{j}|^{2}=1, \sum_{j=0}^{3}z_{j}^{2}=0$. Then, taking $z_{j}=x_{j}+iy_{j}$ $\forall j=0,1,2,3; x_{j},y_{j}\in{\mathbb{R}}$ and utilizing the above restrictions, we get $\sum_{j=0}^{3}x_{j}^{2}=\sum_{j=0}^{3}y_{j}^{2}=\frac{1}{2}$ and $\sum_{j=0}^{3}x_{j}y_{j}=0$. The fruitful implementation of the result of previous theorem and simple algebraic manipulations using these results lead to the fact $N(\rho_{AB})+N(\rho_{AC})+N(\rho_{AD})\leq \frac{1}{4}$.$\blacksquare$   \\\\
\textbf{Theorem 5}: \textit{Let $\rho_{ABCD}$ be any four-qubit pure state belonging to the class $\tau_{min}$. Then $\frac{1}{2}\leq N(\rho_{AB})+N(\rho_{AC})+N(\rho_{AD})\leq \frac{3}{4}$ }.\\\\
\textit{Proof}: Let us consider any four-qubit pure state $\rho_{ABCD}=|\psi\rangle_{ABCD}\langle\psi|$, belonging to the class $\tau_{min}$ i.e., $|\psi\rangle_{ABCD}=\sum_{j=0}^{3}x_{j}u_{j}$, with $\sum_{j=0}^{3}x_{j}^{2}=1 $ and $x_{j}\in\mathbb{R}, j=0,1,2,3 $. Now let us denote the sets of eigenvalues of $\rho_{AB},\rho_{AC}$ and $\rho_{AD}$ as $\{\lambda_{1}^{AB},\lambda_{2}^{AB},\lambda_{3}^{AB} \}$, $\{\lambda_{1}^{AC},\lambda_{2}^{AC},\lambda_{3}^{AC} \}$ and $\{\lambda_{1}^{AD},\lambda_{2}^{AD},\lambda_{3}^{AD} \}$ respectively. Without loss of generality we can consider, $\lambda_{i}^{AB}\geq \lambda_{j}^{AB}\geq \lambda_{k}^{AB}$,$\forall i,j,k=1,2,3$ with $i\neq j\neq k$. This consideration gives natural connections among the eigenvalues of $\rho_{AC}$ and also $\rho_{AD}$. These are $\lambda_{i}^{AC}\geq \lambda_{j}^{AC}\geq \lambda_{k}^{AC}$ and $\lambda_{i}^{AD}\geq \lambda_{j}^{AD}\geq \lambda_{k}^{AD}$, $\forall i,j,k=1,2,3$ with $i\neq j\neq k$. Observing the behavior of the eigenvalues and straight forward derivation quite easily establish the desired result, i.e., $\frac{1}{2}\leq N(\rho_{AB})+N(\rho_{AC})+N(\rho_{AD})\leq \frac{3}{4}$.$\blacksquare$  \\\\

These two results give us information about the monogamous behavior of MIN in two important classes of four-qubit system. Since, $N(\rho_{A|BCD})={\frac{1}{2}}$ in the whole class $\mathcal{A}$, therefore, MIN is monogamous in the subclass $\mathcal{M}$ but it is polygamous in other class  $\tau_{min}$. Further, all the states that are connected to these classes by LU, share the same fate. Four-qubit GHZ state belongs to the class $\tau_{min}$. Hence GHZ and their LU equivalent states are not monogamous w.r.t MIN. Another two states(and obviously their LU equivalent states) $|L\rangle=\frac{1}{\sqrt{3}}(u_{0}+\omega u_{1}+\omega^{2}u_{2})$ where $\omega=e^{\frac{2i\pi}{3}}$ and $|M\rangle= \frac{i}{\sqrt{2}}u_{0}+\frac{1}{\sqrt{6}}(u_{1}+u_{2}+u_{3})$ which maximize the Tsallis $\alpha$-entropy for different regions of $\alpha$ \cite{gour} satisfy the monogamy relation of MIN. On the other hand the four-qubit cluster states satisfies the monogamy relation as their two party reduced density matrices are completely mixed (i.e., MIN is zero). For four-qubit generalized W-states $\rho^{W}=\alpha|1000\rangle+\beta|0100\rangle+\gamma|0010\rangle+\delta|0001\rangle$ where $|\alpha|^{2}+|\beta|^{2}+|\gamma|^{2}+|\delta|^{2}=1$, monogamy relation holds with equality, i.e., it can be easily shown (as in the three-qubit case) that $N(\rho_{AB}^{W})+N(\rho_{AC}^{W})+N(\rho_{AD}^{W})=N(\rho_{A|BCD}^{W})=2|\alpha|^{2}(1-|\alpha|^{2})$. Hence for this type of states nonlocality shows additive property with respect to each party. This result of generalized W- class can be further extended to n-qubit system with the same conclusion.\\

\section{Conclusion}
Thus we have explored the monogamy nature of MIN and found certain classes of states on which MIN shows monogamous nature. Unlike geometric discord,\cite{bruss} MIN can be polygamous for pure states as revealed in some subclasses of three- and four-qubit generic class. On the other hand, W-class seems to satisfy the monogamy relation with equality in n-qubit system. So for the W-class MIN becomes additive in terms of sharing between the parties. The monogamous nature of W-class states w.r.t. MIN in any dimension indicates a distinguishing feature of this class of states. Thus existence of monogamy of this type of correlation put a restriction on the amount of shared nonlocality. Monogamous nature of MIN for these classes of states can be exploited in providing some cryptographic protocol.

{\bf Acknowledgement.} The author A. Sen acknowledges the financial support from University Grants Commission, New Delhi, India.

\end{document}